# A New Dynamic Bandwidth Allocation Protocol with Quality of Service in Ethernet-based Passive Optical Networks

Fu-Tai An, Yu-Li Hsueh, Kyeong Soo Kim, Ian M. White, and Leonid G. Kazovsky
Optical Communication Research Laboratory, Stanford University
350 Serra Mall, Rm. 058, Stanford, CA 94305
USA

## ABSTRACT

Ethernet-based Passive optical network (E-PON) is the key for next generation access networks. It must have the property of high efficiency, low cost, and support quality of service (QoS). We present a novel media access control (MAC) protocol that maximizes network efficiency by using dynamic bandwidth allocation (DBA) algorithm suitable for E-PON. This protocol minimizes packet delay and delay variation for high priority traffic to ensure QoS. Simulation results show excellent network throughput. Simulation results also show low packet delay and packet delay variation for high priority traffic compare with traditional MAC protocol of E-PON. When the network performs ranging, this protocol ensures zero interruption of high priority traffic, such as audio or video applications.

## 1. INTRODUCTION

Due to the development of optical communications, the Internet backbone has been using optical fibers for long-haul, high throughput transmission. However, end users still cannot enjoy the luxury of high bandwidth and good signal quality that optical communication provides. This is due to the bottleneck of access networks in terms of technology development and network architecture design. Many researches now are focusing on optical access networks [1]. Passive Optical Networks (PON) is the most promising candidate for optical access network due to its simplicity.

Minimal protocol overhead and low-cost nature of Ethernet makes it appealing to be chosen as the packet format of PON. Many researches are now focusing on Ethernet-based PON (E-PON) [2]. However, there are two major challenges for E-PON. First is the media access control (MAC) protocol design to ensure high throughput, since traditional CSMA/CD MAC protocol for Ethernet cause severe performance degradation in E-PON due to the large collision domain [3]. Second is meeting Quality of Service (QoS) constraint. Ethernet does not intrinsically support QoS, but supporting QoS is essential for E-PON to accommodate various traffic types. In this paper, we propose a novel MAC protocol employing dynamic bandwidth allocation (DBA) to maximize network efficiency and ensuring QoS. Network efficiency is optimized by means of hybrid slot size/rate algorithm to minimize guard time between packets. High priority traffic is always assigned in the fixed location of the frame to minimize delay variation. Also, when a new node joins the network and initial handshaking is ongoing, high priority packets with stringent delay and delay variation constraint will not be disturbed. We present the simulation results on network throughput, average packet delay, and packet delay variation to demonstrate validity our ideas.

## 2. PASSIVE OPTICAL NETWORK STRUCTURE

Fig. 1 shows the typical architecture of PON. We believe that this tree structure with single-wavelength TDM is an attractive solution. In traditional TDM solution, the Optical Line Terminal (OLT) assigns fixed time slots in upstream traffic for each Optical Network Unit (ONU). Since ONUs have different distances to OLT, ranging has to be performed such that the gaps between slots are minimal.

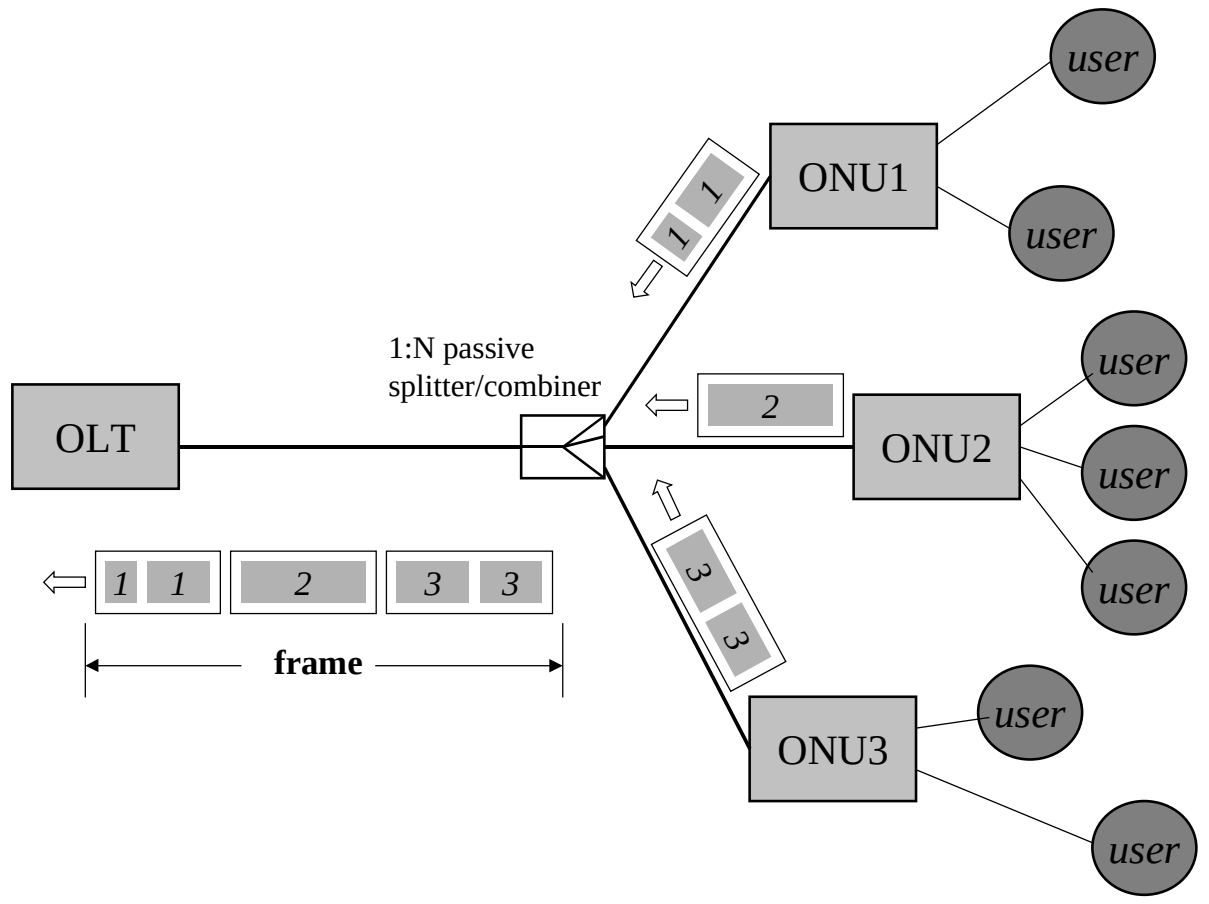

**Fig. 1**. Architecture for PON with DBA for upstream traffic

Since the fiber link is a shared media for all ONUs, maintaining high network efficiency is very important, especially for upstream traffic. We may employ conventional fixed-slot TDM or even statistical multiplexing to solve the problem of packet collision. However, due to the burstness of traffic, the unused time slots prohibit the network from reaching its optimum efficiency. Hence, MAC protocol that uses *dynamic bandwidth allocation* (DBA) algorithm to dynamically perform traffic management for upstream traffic is necessary. Within each frame, each ONU sends a request or reports its queue size to OLT. Since OLT has global information, it allocates appropriate bandwidth to each ONU on the fly [4].

One important parameter in this network design is guard time (GT) between packets from different ONUs for upstream traffic. The guard time consists of several factors: laser switching delay, ranging inaccuracy, and clock-recovery preamble. Because the ONU's laser must be turned off completely when it is not transmitting, the lower bound of GT is laser's switching time, which is about 10ns. Taking into account various practical issues, GT of sub-μs is reasonable.

# 3. MEDIA ACCESS CONTROL PROTOCOL DESIGN

## 3.1 Protocol Overview

Conventional E-PON uses a slot-size based DBA algorithm. That is, within the grant cycle (frame), the slots assigned to ONUs are proportional to the traffic demand from ONUs. Packets of all priorities are put in the same slot. This is mainly due to the reason that there is substantially a large guard time between traffic from different ONUs. By using this slot-size based DBA, the number of occurrences of guard time is equal to the number of ONUs. Due to the burstness of traffic and packet length variation, the starting point of each slot is push-pulled from frame to frame. Although the repetition of frame ensures the packets' delay, delay variation of packet is not under control. Worst-case packet delay variation (PDV) can be equal to time duration of one frame. This is shown in Fig. 2(a).

In our design, we divide a frame into two parts, one is the steady part that has *N* time slots for high priority traffic, corresponding to *N* ONUs, and the other is the dynamic part that is basically one giant slot for low priority traffic, such as best-effort (BE) traffic, shared by ONUs. Since the amount of high priority traffic flow from ONUs is determined at the time users subscribe for the service, the steady part of the frame can be approximated by constant flows. Usually, the high priority traffic requires much less bandwidth, but is very sensitive to delay and delay variation. For the low priority class traffic, there is no requirement on any QoS parameters; therefore, optimizing throughput is the goal. We name this protocol *hybrid slot-size/rate DBA* protocol. This is demonstrated in Fig. 2(b). In order to deliver the concept clearly, we assume there are only two service classes, HP and BE, in this case.

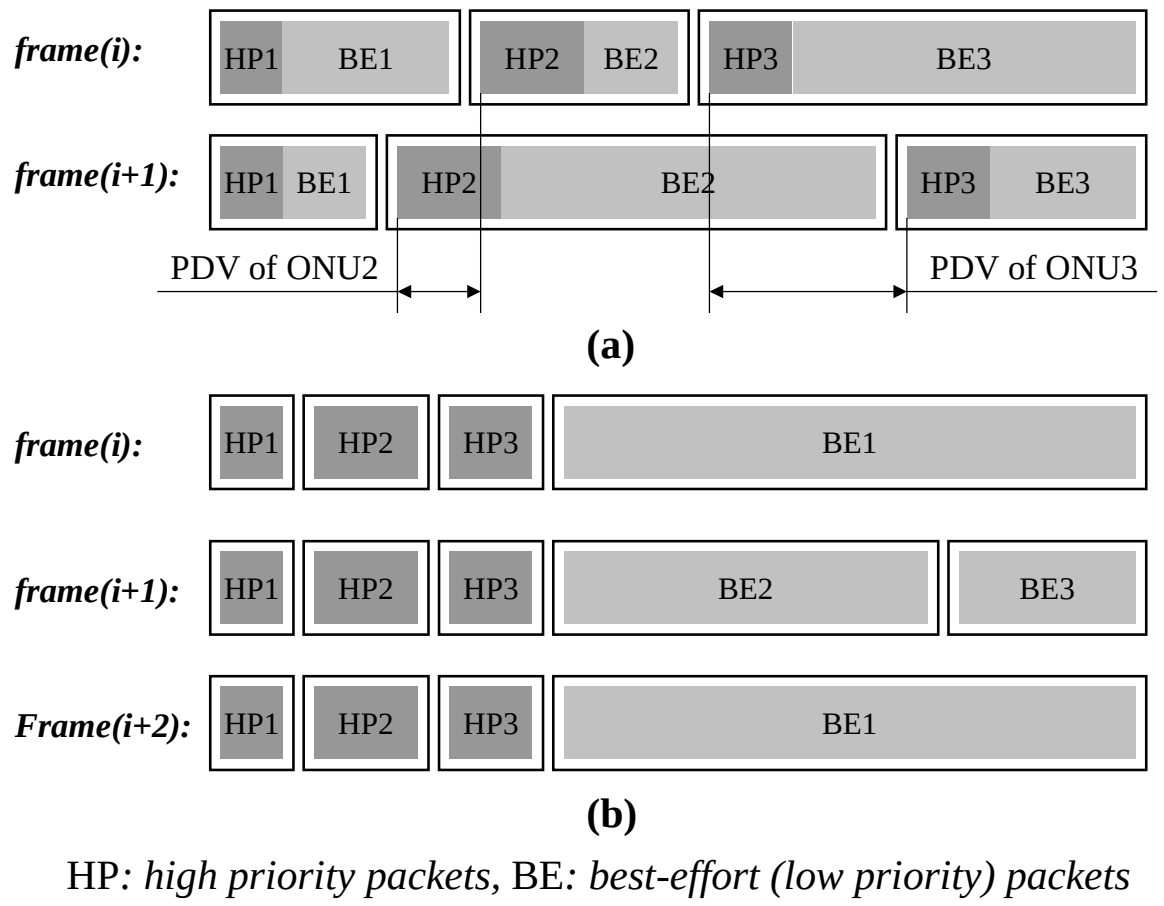

**Fig. 2.** (a) Packet delay variation in conventional slot-size based DBA; (b) hybrid slot-size/rate (HSSR) DBA

When the network load is low, OLT can order several ONUs to transmit their low priority data in the dynamic part of a frame. Since network is not fully utilized, the guard time between packets of different ONUs has minor impact on network performance. When network load is very high, OLT dedicate the dynamic part in one frame to particular ONU, and assign the dynamic part in other frames to another ONU. In other words, an ONU may solely occupy the dynamic part in a frame, and we can save the overhead due to guard time.

The delay constraint set by real time application dictates the frame repetition rate. For example, in telephony, a user starts to notice delay when it exceeds 150mS. This includes the propagation delay over the Internet. According to T1.508, for access network to avoid echo-cancellation circuitry, the data has to be less than 5ms. Various sources will affect the delay in the network, such as queuing delay, jitter buffer, etc. In general, frame repetition rate is on the order of ms.

## 3.2 Packet Delay Variation Minimization

Unlike the conventional slot-size based DBA scheme where traffic of all classes from single ONU is mixed together, we effectively separate traffic of different service class in different part of the frame. Effectively the protocol establishes circuit-like channels for the high priority upstream traffic from each ONU to OLT. Therefore, the relative timing for each of the steady slots to the starting of frame is always fixed. This means that delay variation due to slot push-pull is effectively zero. The dedicated slot in a frame ensures the throughput. We can easily and precisely meet QoS requirement for real-time applications.

For low priority traffic, its behavior differs as network load varies. If the network traffic load is light, and we can allocate transmission for multiple or all ONU in the dynamic part, the delay variation of these low priority packets is generally the same as that in the conventional slot-size based DBA. When network load is heavy, each ONU may be granted to transmit low priority data every N frames. Delay of packets of low service class is significantly prolonged. However, there is no constraint on the delay for low priority class at all. The only penalty is that ONU needs longer queue to hold the data. The benefit of bypassing frames for low priority traffic is saving overhead due to guard time.

### 3.3 Packet Reordering

Packet class reordering is necessary to guarantee fairness and increase network throughput. When a user pumps up the high priority traffic that exceeds the rate that was subscribed, ONU will redirect the surplus traffic to the queue of lower service class. In other words, inside ONU, there are multiple class-based queues corresponding to different service classes. ONU is bookkeeping on data ingress rate to make sure that users do not take advantage of occupying more bandwidth than they subscribed. The idea of packet class reordering in ONU is demonstrated in Fig 3. Here only two classes are shown, but the idea can be extended to multiple service classes.

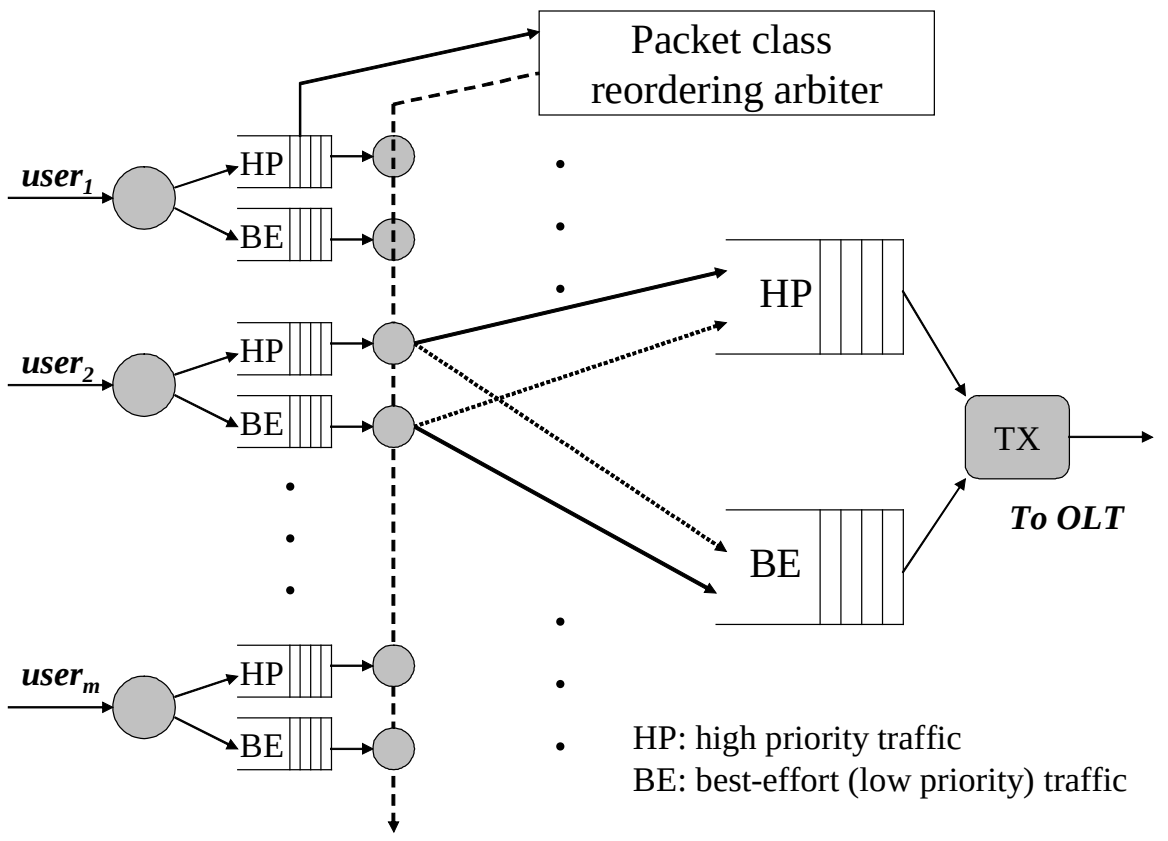

**Fig. 3.** Packet class reordering in ONU

Besides lowering traffic class to ensure fairness, it is also possible to move traffic class up to increase efficiency. High priority traffic, such as telephony or video conferencing, is a traffic flow set up for minutes or hours. However, the flow may not always exist. In our protocol, we reserve the bandwidth for fully utilized high priority traffic. The steady slots are wasted if they are not used for sending data. Therefore, lower priority class traffic can be promoted to fulfill these slots. There are two places this process can occur. First, the user's application can calculate the rate for different services. The user can change the QoS tag of packets to achieve the goal. Second, the ONU is performing bookkeeping of its own. It will move some low priority data to high priority queue if the user is not fully utilizing the bandwidth and if the packet can fit into the steady slots.

### 3.4 Quasi-Non-Intrusive Ranging

For certain time duration, such as ten seconds, OLT has to detect if there is a new ONU to join this network. If the answer is positive, OLT has to perform ranging to determine the distance of that ONU, and necessary initialization as well. OLT will send out a token and listen for the reply from new ONU. Due to the uncertainty of the distance between OLT and ONU, and because the link is a shared media, the reply of the newly joined ONU is likely to collide with data from other existing ONUs. This may cause data retransmission and long ranging time, and is undesirable. The straightforward solution is for OLT to send out "QUIET" signal to all ONU. The reply from new ONU then can be detected easily. This is called *intrusive ranging*. Ranging may iterate several rounds to acquire bit-level timing accuracy. The interruption of data may cause glitches or loss of connection or real-time applications. Non-intrusive ranging is very desirable to keep high priority traffic intact. Research has proposed an analog method [5]; however, it requires extra RF circuitry. If we take a close look of the packet characteristics, only high priority traffic is sensitive to interruption from ranging. BE traffic is immune to interruption.

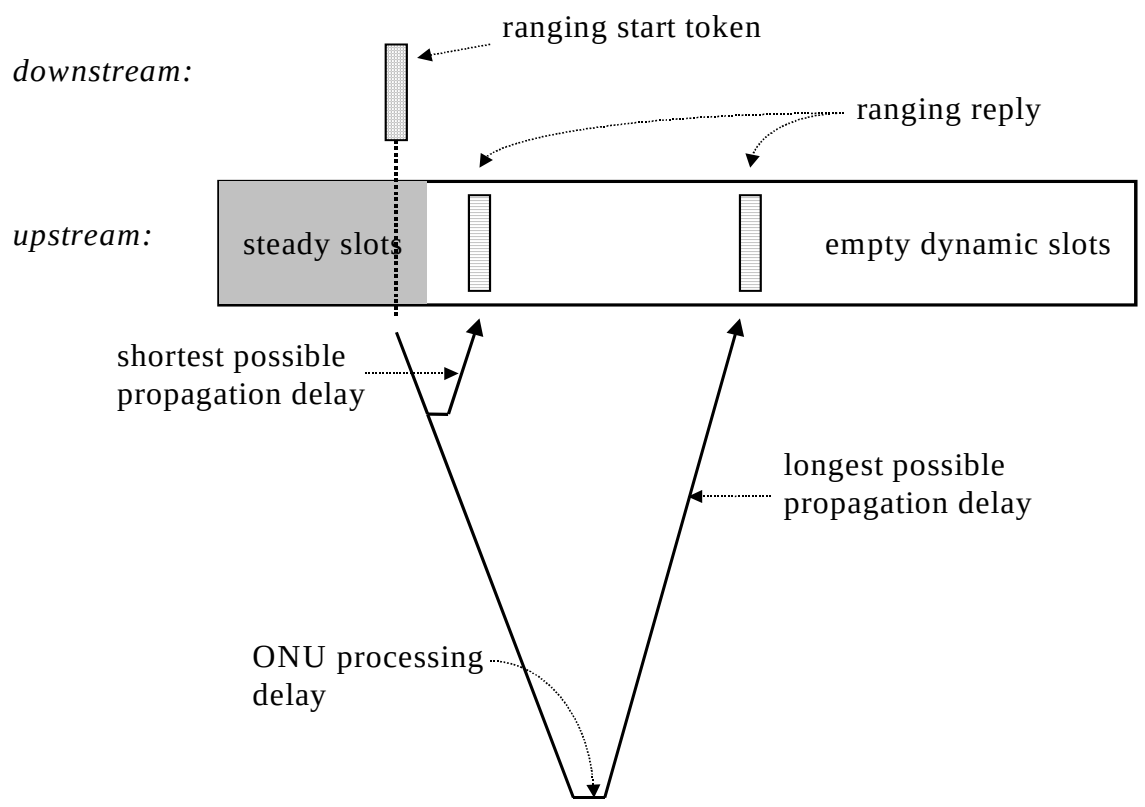

**Fig. 4.** Quasi-non-intrusive ranging

In our scheme, when OLT performs ranging, *only BE traffic* is shut down temporarily. Since we know the distance range, by carefully placing the ranging token, the ranging reply from new ONU is guaranteed to fall inside the empty BE part of the frame. This means that real-time applications are not perturbed. Fig. 4 demonstrates this idea. Here we assume that the fiber length from OLT to ONU is within 2km to 20km. Therefore, the propagation delay is in the range of 10µs to 100µs. The total response time of ranging process is much less than frame size, 1ms. The protocol guarantees OLT obtain ranging reply within the current frame.

### 3.5 Hybrid Slot Size/Rate (HSSR) Algorithm

Since frame size is much longer that propagation time, we can assume that upstream frame and downstream frame are aligned. Within each frame, an ONU generates grant request by reporting its queue length associated with each class in terms of number of bytes to OLT. No further detailed information shall be reported. When the grant request arrives to OLT, it updates a table that OLT keeps. This table contains ONU number, ONU queue length of each class, and a counter. This counter is to guarantee fairness. If certain ONU does not have the right to transmit low priority packets in current frame, OLT increments the counter; on the other hand, if ONU can transmit low priority packets in current frame, OLT resets the counter. The decision of bandwidth assignment is based on the queue length weighted by the counter value. Therefore, OLT ensures low-load ONUs have the right to transmit data. OLT sends grant reply back to ONUs each frame in downstream data. The grant reply indicates the amount of bandwidth the ONU is assigned for the next upstream frame.

# 4. SIMULATION RESULTS AND DISCUSSION

We model the PON as a tree structure, where there is one OLT at the root, and 8 to 32 ONUs at the node connected to the OLT. There are also several users connected to ONU that act as traffic source. We assume the frame size is 1ms, and the possible guard time is from 20ns to 1μs. The traffic distribution is assumed to be Poisson process. In the simulation, the packet size distribution is as [6], where the packet size distribution is discreet with popular packet sizes of 40, 552, and 1500 bytes.

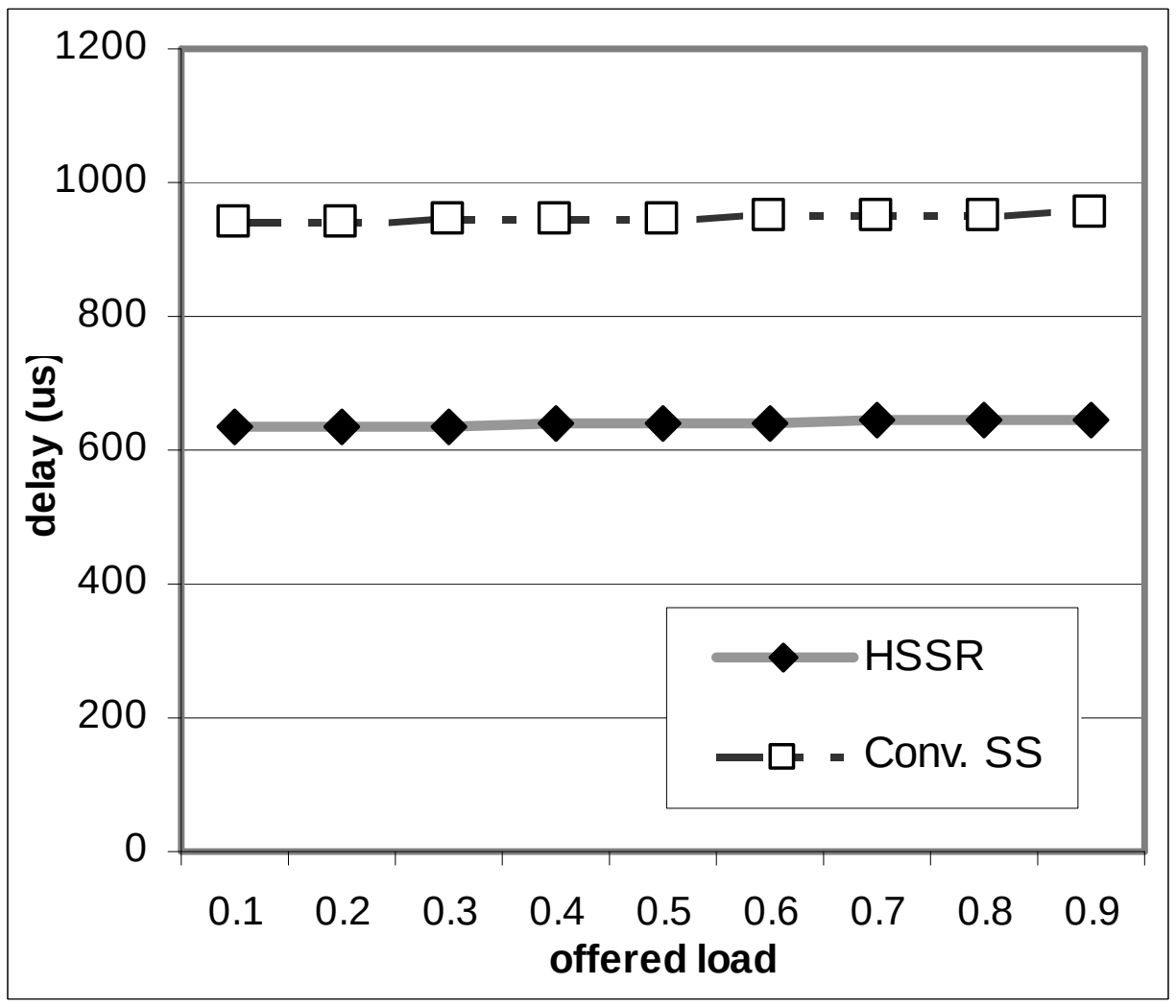

**Fig. 5.** Packet delay versus network load

Fig. 5 shows the simulation result of average packet delay. It focuses on the average packet delay for high priority traffic. Under *hybrid slot-size/rate (HSSR)* DBA scheme, the average packet delay is always about 640us regardless of offered load. 500us out of 640us is the half of the frame size, which is also the average time for a head-of-line packet in the queue to be sent to OLT. 140us out of 640us is due to queuing delay. The high priority traffic of conventional slot-size based DBA has much longer packet delay, which is about 950us. Compare with that of HSSR scheme, the extra 310us is due to extra delay caused by the uncertain positioning of packets inside frame. Simulation results also show that BE traffic belongs to HSSR has the tendency of having extra 100us delay compare with that of conventional SS scheme. This is not an issue since delay is not critical to BE traffic.

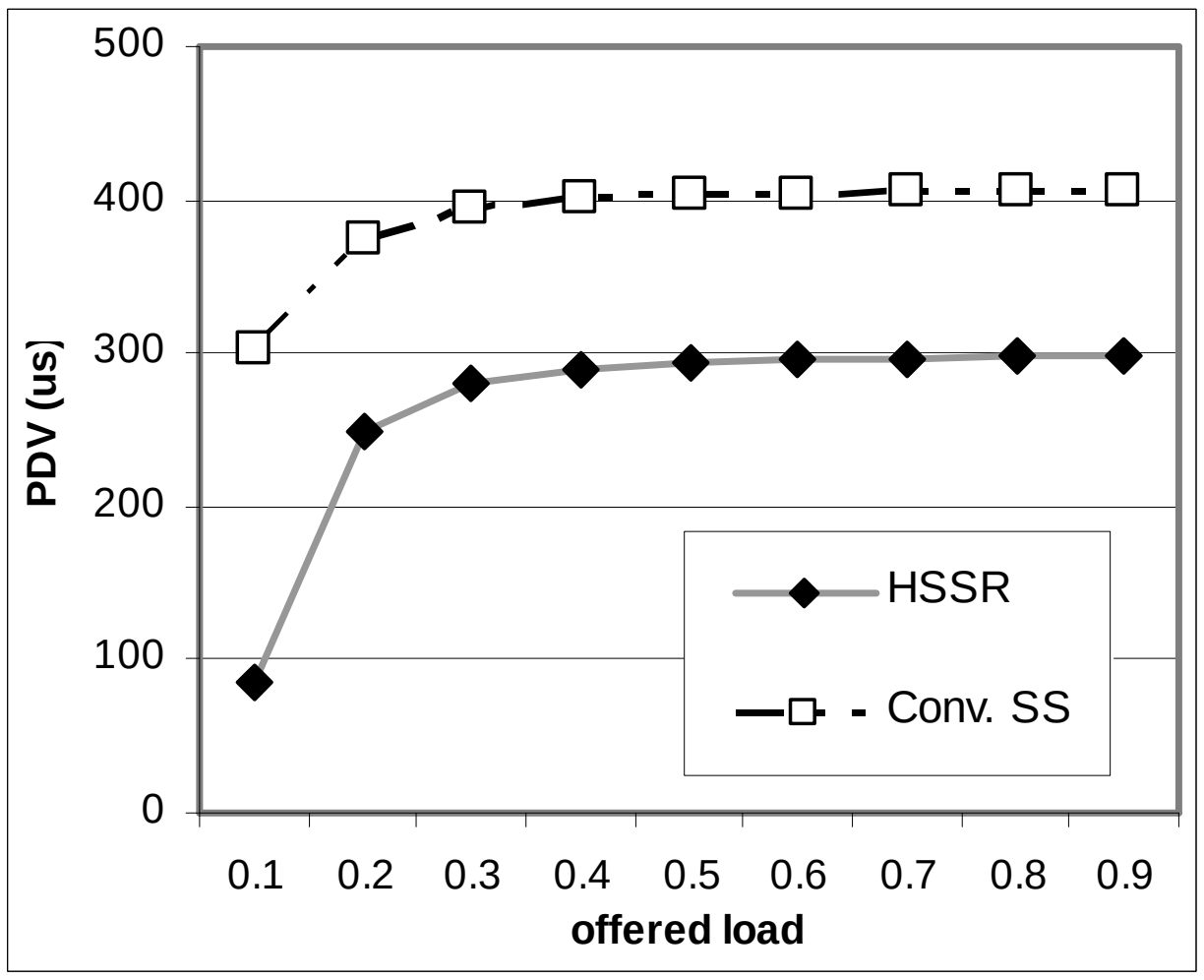

**Fig. 6.** Packet Delay Variation

Fig. 6 shows the packet delay variation (PDV) of high priority traffic for both HSSR algorithm and conventional slot-size (SS) based algorithm. Here we use the standard deviation of packet delay to represent packet delay variation. The PVD of high priority packets in HSSR is always less than 300us. If is about 100us less than that of high priority packets in the conventional SS scheme. The delay variation is caused only by queuing, not by DBA algorithm for E-PON itself. This is because we fix the bandwidth and slot location for high priority traffic in HSSR algorithm.

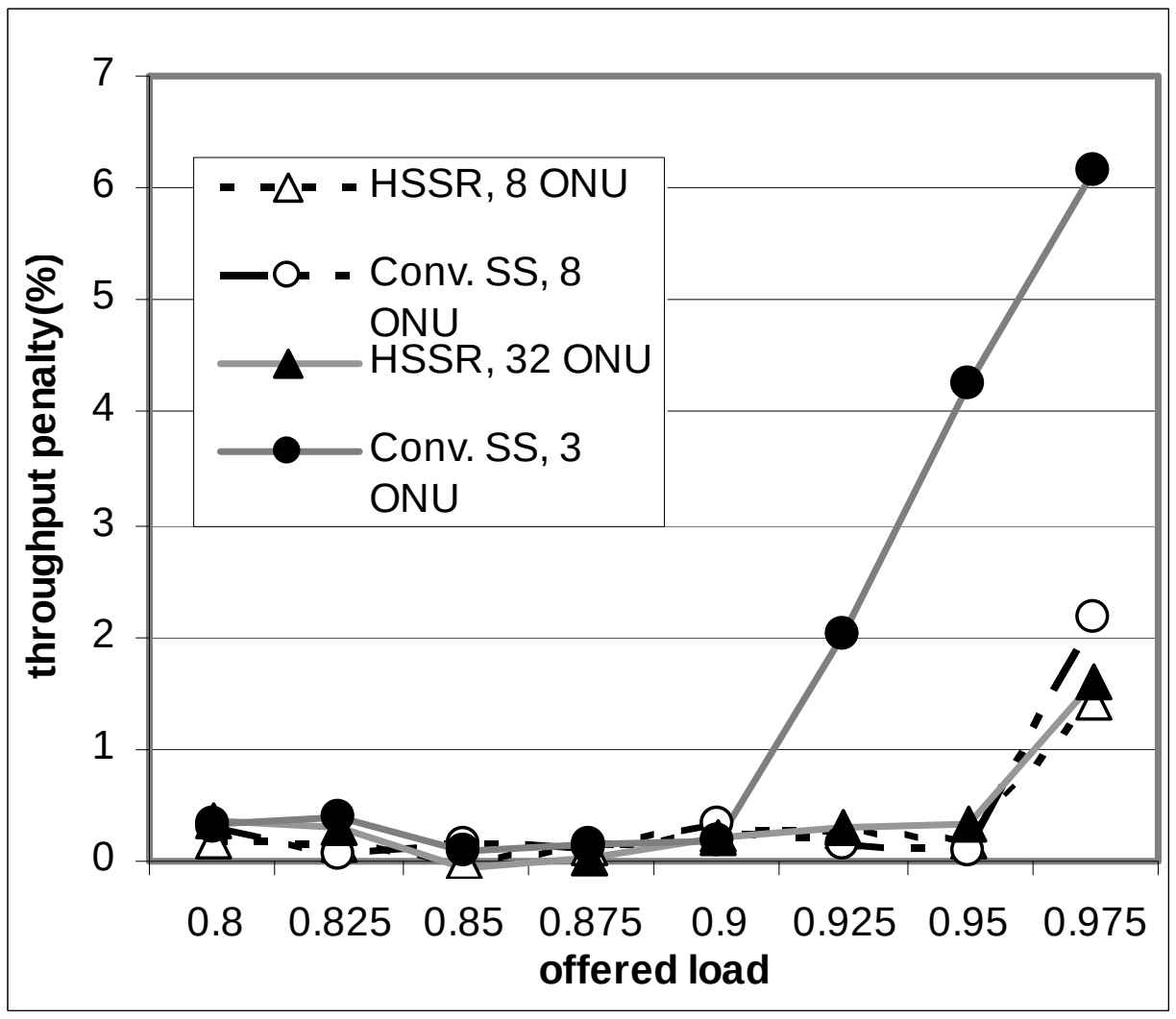

**Fig. 7.** Throughput penalty of low priority (BE) traffic

Although there is no constraint on delay or delay variation for BE traffic, it is important to maximize the throughput. Fig. 7 shows the throughput penalty of low priority BE traffic. The vertical axis means the ratio of transmitted BE traffic to maximum possible BE traffic, and the horizontal axis is the offered load. In this simulation we assume that the BE traffic is 700Mb/s, while total bandwidth for all traffic is 1Gb/s. Also, we assume the guard time is 0.1μs, and frame size is 1ms. In HSSR algorithm, when load is close to 1, there is a penalty that is less than 2%. This is due to the overhead of extra guard time for BE slots. On the other hand, the SS algorithm has much larger penalty compared with HSSR. When the number of ONU is 32, there can be over 6% reduction in transmission rate for conventional slot-size based algorithm. The cause of the phenomena is that when the load is high and many ONUs are fighting for the resource, each of them will get only a small slot. For this small slot, ONU will be challenged to fully utilize the slot. Since the packet size is a variable from 40 bytes to 1,500 bytes, even with packet look-ahead in the queue, it is still very hard for an ONU to schedule the packets such that the unused portion of the assigned slot is minor.

We have noticed that the conventional slot-size (SS) based algorithm can be modified to improve throughput of BE traffic. One way is for ONU to convey detailed information of its queue to OLT and let OLT perform global optimization of scheduling. However, each ONU supports multiple users and the detailed dynamic traffic profile can consume significant portion of upstream connection. The other way is to postpone BE transmission of several ONUs to following frames. This is similar to HSSR, but without separating high priority and low priority packets. Doing so will make the PDV of high priority traffic even worse.

## 5. CONCLUSION

In this paper we have discussed the E-PON architecture and the dynamic bandwidth allocation scheme for optimizing network efficiency. We have also addressed issues of implementing QoS in E-PON.

We have proposed the hybrid slot-size/rate DBA MAC protocol. This protocol is simple and scalable for next generation E-PON. This protocol ensures that high priority traffic will have short delay and minimal packet delay variation. The size of elastic buffer, which is used to mitigate delay and delay variation, in the network can therefore be dramatically reduced. The protocol assigns a large slot size for low priority, such as best effort, traffic. This maximizes network throughput when offered load is high. The fairness for low priority is guaranteed by using queue length as weighting factor. The protocol also provides quasi-non-intrusive ranging. Ranging process will not disturb the high priority traffic.

We have verified this concept using a simulation that compares HSSR DBA and conventional slot-size based DBA algorithm. Our protocol shows excellent network efficiency. Also, packet delay and delay variation parameters are much better then conventional slot-size based DBA MAC. Packet delay variation comes solely from queuing but not traffic scheduling of the protocol.

## 6. ACKNOWLEDGEMENT

The authors would like to thank Dr. Chung-Li Lu for fruitful discussion.

## REFERENCES

[1] C. Pantjiaros, C. Combes, R. Van Wolfswinkel, et. al., Broadband Services Delivery over an ATM PON FTTx System, *Electrotechnical Conference*, 2000, pp. 225-229

[2] IEEE 802.3ah EFM Task Force P2MP (EPON) Baseline Proposals, http://www.ieee802.org/3/efm/ baseline/p2mpbaseline.html

[3] G. Kramer et al., “EPON scheduling protocol requirements,” IEEE 802.3ah EFM Task Force Meeting, Jun. 2002.

[4] G. Kramer and G. Pesavento, “Ethernet Passive Optical Network (EPON): Building a Next-Generation Optical Access Network”, IEEE Communications Magazine, Feb 2002, vol. 40, no. 2, pp. 66-73

[5] P. Van Hejiningen, T. Mosch, et. al., Out of band ranging method for ATM over PON access systems, *Proc. ECOC* 1994, pp.271-274

[6] WAN packet size distribution (http://www.nlanr.net/NA/Learn/packetsizes.html)